\documentclass[10pt, aps, pra, twocolumn, floatfix]{revtex4}
\usepackage{amsmath, amssymb, mathtools, array, color, bbm}
\usepackage{hyperref}
\hypersetup{colorlinks=true, linkcolor=blue, citecolor=blue}
\usepackage{graphicx}
\usepackage{times}

\newcommand{\bra}[1]{\langle{#1}|}
\newcommand{\ket}[1]{|{#1}\rangle}

\newcommand{\ex}[1]{\langle {#1}\rangle}

\newcommand{\trm}{\textrm}

\newcommand{\pd}{\phantom{\dag}}

\newcommand{\nn}{\nonumber}

\newcommand{\fd}{\mathcal{D}}

\begin{document}


\title{Quantum synchronization of two Van der Pol oscillators}

\author{Stefan~Walter}
\author{Andreas~Nunnenkamp}
\author{Christoph~Bruder}
\affiliation{Department of Physics, University of Basel, Klingelbergstrasse 82, CH-4056 Basel, Switzerland}

\date{\today}

\date{\today}


\begin{abstract}
We study synchronization of two dissipatively coupled Van der Pol oscillators in the quantum regime. Due to quantum noise strict frequency locking is absent and is replaced by a crossover from weak to strong frequency entrainment. We discuss the differences to the behavior of one quantum Van der Pol oscillator subject to an external drive. Moreover, we describe a possible experimental realization of two coupled quantum van der Pol oscillators in an optomechanical setting.
\end{abstract}

\maketitle

\section{Introduction}\label{sec:intro}

Synchronization is a ubiquitous phenomenon occurring in a variety of physical systems consisting of so-called self-sustained (or limit-cycle) oscillators~\cite{Pikovsky2001}. The key difference between a self-sustained oscillator and a coherently driven system is that the phase of a self-sustained oscillator is not fixed by the phase of the external drive, so it is free to synchronize.

Different scenarios of synchronization can be considered. The simplest setting is a limit-cycle oscillator coupled to an external harmonic drive. With increasing drive strength the frequency of the limit-cycle oscillator is pulled to the frequency of the drive. If the two frequencies lock together, oscillator and drive have a fixed relative phase. A second possible setting consists of two coupled limit-cycle oscillators. A third well-studied case is the limit of infinitely many coupled limit-cycle oscillators with random frequencies, for example, the Kuramoto model~\cite{Kuramoto1984, Acebron2005} showing non-equilibrium phase transitions from unsynchronized to synchronized states.

Synchronization in \emph{classical} nonlinear dynamical systems has been widely studied and is a well-understood phenomenon. However, the subject of synchronization in the \emph{quantum} regime has been considered only recently~\cite{Shepelyansky2006, Haenggi2006, Ludwig2012, Lee2012, Mari2013, Hriscu2013, Lee2013, Holland2013, Walter2014, Lee2014}. In the classical case the paradigmatic example for a self-sustained oscillator is the Van der Pol (VdP) oscillator. This is a nonlinear dynamical system with two kinds of damping terms: negative damping which makes the rest position unstable and nonlinear damping which in turn limits the motion of the oscillator to a finite amplitude. Synchronization in the quantum version of the VdP oscillator subject to an external harmonic drive has recently been investigated~\cite{Lee2013, Walter2014}. It was shown that phase locking~\cite{Lee2013} and frequency entrainment~\cite{Walter2014} survive in the quantum regime, i.e.,~where the system is close to its quantum ground-state.

In this paper, we extend our previous study~\cite{Walter2014} to the case of two coupled quantum VdP oscillators~\cite{Lee2014} where we mainly focus on frequency entrainment. We use the quantum-mechanical power spectrum to characterize the frequencies of two dissipatively coupled quantum VdP oscillators and find that genuine frequency locking is absent in the quantum regime due to quantum noise. Compared to classical (noiseless) coupled VdP oscillators the synchronization region is reduced. We find regions of strong and weak frequency entrainment between the quantum VdP oscillators depending on the coupling strength. For detuned, but otherwise identical limit-cycle oscillators the entrainment behavior as a function of the coupling is different from the entrainment behavior of an externally driven quantum VdP~\cite{Walter2014}. For non-identical quantum VdP we show that the oscillator with the smaller limit cycle changes its frequency faster in response to the coupling.

A particularly promising architecture to study quantum synchronization are opto- and nanomechanical systems~\cite{AKM2013}. These systems offer the possibility to engineer nonlinearities and dissipation and thus to control their dynamics in the quantum regime. A number of experiments on synchronization in the classical regime have recently been reported with nanomechanical systems~\cite{McEuen2012, Bagheri2013, Roukes2014}.

The paper is organized as follows. In Sec.~\ref{sec:vdpCoupCl} we write down the classical equations of motion for coupled VdP oscillators in the classical domain and then introduce the model of two dissipatively coupled quantum VdP oscillators in Sec.~\ref{sec:vdpCoupQm}. We investigate their synchronization in terms of phase locking in Sec.~\ref{sec:steadystate} and study frequency entrainment in Sec.~\ref{sec:entrainment}. In Sec.~\ref{sec:optomech} we discuss a possible realization of our model in the framework of cavity optomechanical systems and conclude in Sec.~\ref{sec:conclusion}.

\section{Two coupled classical Van der Pol oscillators}\label{sec:vdpCoupCl}

Let us begin by writing down the equations of motion for two coupled VdP oscillators~\cite{Pikovsky2001}
\begin{align}
	\dot{\alpha}_{1} &= -i \omega_{1} \alpha_{1} + \frac{\gamma_{1}^{(1)}}{2} \alpha_{1} - \gamma_{2}^{(1)} |\alpha_{1}|^{2} \alpha_{1} \nn \\
						&+ \left( \frac{D}{2} +i R \right) \left( \alpha_{2} - \alpha_{1} \right) \label{eqn:vdpC3a} \, , \\
	\dot{\alpha}_{2} &= -i \omega_{2} \alpha_{2} + \frac{\gamma_{1}^{(2)}}{2} \alpha_{2} - \gamma_{2}^{(2)} |\alpha_{2}|^{2} \alpha_{2} \nn \\
						&+ \left( \frac{D}{2} +i R \right) \left( \alpha_{1} - \alpha_{2} \right) \label{eqn:vdpC3b} \, ,
\end{align}
where $\gamma_{1}^{(i)}$ and $\gamma_{2}^{(i)}$ describe negative and nonlinear damping of the $i$-th oscillator, respectively. Negative damping makes rest an unstable fixed point while nonlinear damping limits the motion of the oscillator to a finite amplitude. The stable steady-state solution to Eqs.~(\ref{eqn:vdpC3a}) and (\ref{eqn:vdpC3b}) for $R=D=0$ is a limit cycle with amplitude $|\alpha_{i}| = \bar{r}_{i} = \sqrt{\gamma_{1}^{(i)}/2\gamma_{2}^{(i)}}$, i.e., the ratio of $\gamma_{1}^{(i)}/\gamma_{2}^{(i)}$ determines the size of the limit cycle. The oscillators have natural (intrinsic) frequencies $\omega_1$ and $\omega_2$, respectively. The term proportional to the coupling constant $D$ ($R$) describes dissipative (reactive) coupling.

Further insight can be gained if we separate the complex amplitude $\alpha_{i}$ into modulus and phase, $\alpha_{i} = r_{i} e^{i\phi_{i}}$,
\begin{align}
\dot{r}_{1} &= \left( \frac{\gamma_{1}^{(1)}}{2} - \gamma_{2}^{(1)} r_{1}^{2} \right) r_{1} + \frac{D}{2} \left( r_{2} \cos\theta - r_{1} \right) \nn \\ &-R r_{2} \sin\theta \label{eqn:vdpC4a} \, , \\
\dot{r}_{2} &= \left( \frac{\gamma_{1}^{(2)}}{2} - \gamma_{2}^{(2)} r_{2}^{2} \right) r_{2} + \frac{D}{2} \left( r_{1} \cos\theta - r_{2} \right) \nn \\ &-R r_{1} \sin\theta \label{eqn:vdpC4b} \, , \\
\dot{\theta} &= - \left( \omega_{2} - \omega_{1} \right) + R \left( \frac{r_{1}}{r_{2}} - \frac{r_{2}}{r_{1}} \right) \cos\theta \nn \\ &- \frac{D}{2} \left( \frac{r_{1}}{r_{2}} + \frac{r_{2}}{r_{1}} \right) \sin\theta \label{eqn:vdpC4c} \,.
\end{align}
Eqs.~(\ref{eqn:vdpC4a})-(\ref{eqn:vdpC4c}) depend only on the relative phase $\theta = \phi_{2} - \phi_{1}$.

Synchronization due to dissipative and reactive coupling is qualitatively different~\cite{Pikovsky2001, Kuznetsov2009}. For example, reactive coupling leads to bistability between in-phase and out-of-phase synchronization that would complicate our analysis. In the following we will therefore focus on dissipative coupling and set $R=0$. At small coupling, the dynamics of $r_i$ can be neglected, and Eq.~(\ref{eqn:vdpC4c}) is the Adler equation~\cite{Adler1946}
\begin{align}\label{eqn:adler}
	\dot{\theta} &= - \Delta - \frac{D A}{2} \sin\theta \, ,
\end{align}
where $A = \left( \bar{r}_{1} / \bar{r}_{2} + \bar{r}_{2} / \bar{r}_{1} \right)$ and 
\begin{align}
	\Delta = \omega_{2} - \omega_{1}
\end{align}
is the detuning. Frequency locking in this classical noiseless system will occur for $D \ge 2 |\Delta|/A$.

The Adler equation (\ref{eqn:adler}) has the same form as the one for a VdP oscillator subject to an external harmonic drive. The coupling strength $D$ plays the role of the driving strength, and the detuning $\Delta$ takes the role of the detuning between the frequency of the VdP oscillator and the external drive.

\section{Two coupled quantum Van der Pol oscillators}\label{sec:vdpCoupQm}

Two uncoupled quantum VdP oscillators are described by the master equation (with $\hbar = 1$)~\cite{Lee2013, Walter2014}
\begin{align}\label{eqn:ma0}
	\dot{\rho} = -i \left[ H_{0}, \rho \right] + \sum_{i=1,2} \gamma_{1}^{(i)} \fd[a_{i}^{\dag}] \rho + \gamma_{2}^{(i)} \fd[a_{i}^{2}] \rho \equiv \mathcal{L}_{0} \rho
\end{align}
where $H_{0} = \sum_{i=1,2} \omega_{i}^{\pd} a_{i}^{\dag} a_{i}^{\pd}$ and the coefficients $\gamma_{1}^{(i)}$ and $\gamma_{2}^{(i)}$ describe negative and nonlinear damping of the $i$-th oscillator, respectively. $a_{i}$ are bosonic annihilation operators, and $\mathcal{D}$ are the usual Lindblad dissipators, $\mathcal{D}[O]\rho = O \rho O^{\dag} - \left\{ O^{\dag} O, \rho \right\}/2$. The classical equation of motion for the complex variable $\alpha_{i} = \ex{a_{i}}$ for the Liouvillian $\mathcal{L}_{0}$ are given by Eqs.~(\ref{eqn:vdpC3a}) and (\ref{eqn:vdpC3b}) with $R=D=0$. The ratio $\gamma_{2}^{(i)}/\gamma_{1}^{(i)}$ in the quantum case determines the importance of the discrete level structure: with increasing nonlinear damping rate $\gamma_{2}^{(i)}$ the quantum VdP oscillator is more and more restricted to the lowest Fock states. In the limit $\gamma_{2}^{(i)}/\gamma_{1}^{(i)} \to \infty$ only the two lowest Fock states $\ket{0}$ and $\ket{1}$ are occupied.

Dissipative coupling between the two VdP oscillators is described by the dissipator, $D \fd[a_{1} - a_{2}] \rho$, where $D>0$ is the dissipative coupling. Equations~(\ref{eqn:vdpC3a}) and~(\ref{eqn:vdpC3b}) with $R=0$ are the classical equations of motion for the complex amplitude $\alpha_{i}=\ex{a_{i}}$ corresponding to the full quantum master equation~\cite{Lee2014},
\begin{align}\label{eqn:maD}
	\dot{\rho} = \mathcal{L}_{0} \rho + D \fd[a_{1} - a_{2}] \rho \,.
\end{align}
\begin{figure*}
\begin{center}
\begin{tabular}{c}
	\includegraphics[width=0.7\textwidth]{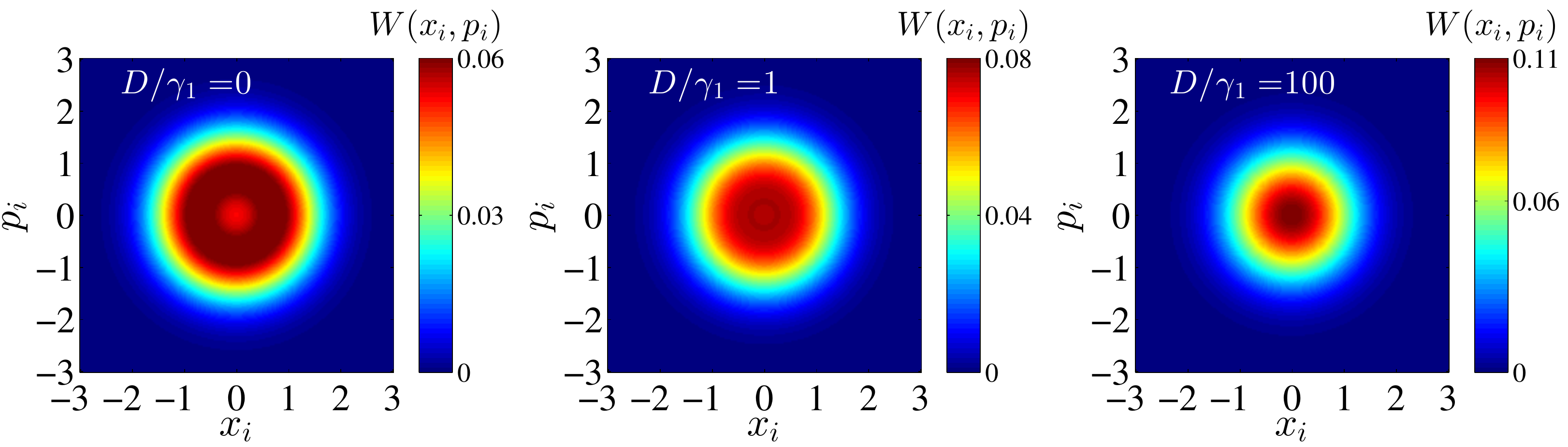}\\
	\hline
	\includegraphics[width=0.7\textwidth]{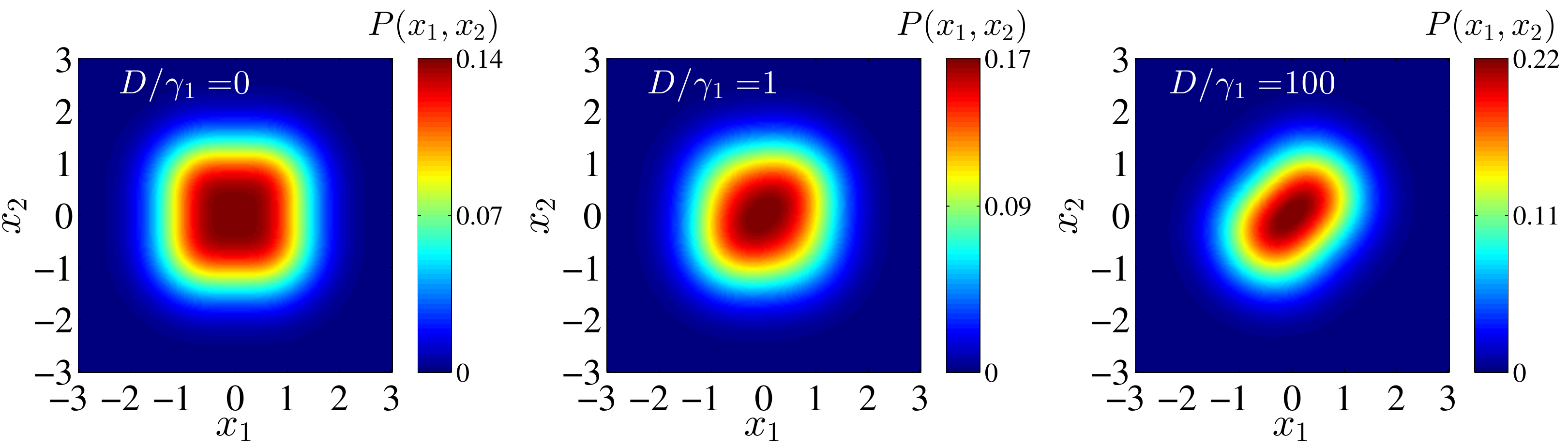}\\
	\hline
	\includegraphics[width=0.7\textwidth]{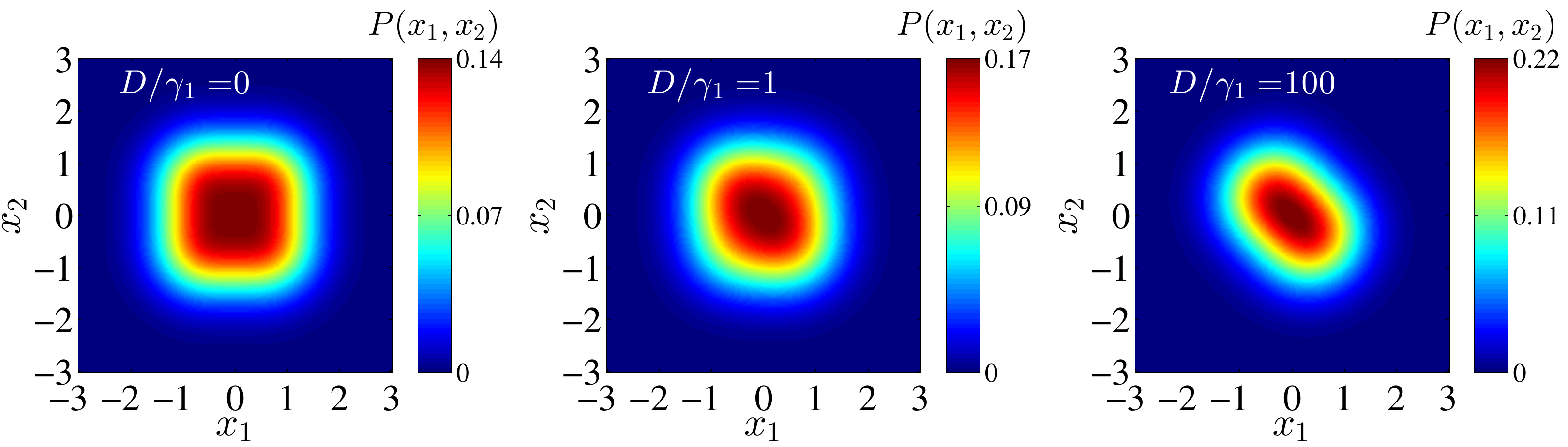}
\end{tabular}
\end{center}
\caption{Wigner function $W(x_{1}, p_{1}) = W(x_{2}, p_{2})$ (first row) and joint-probability distribution $P(x_{1}, x_{2})$ (second and third row) for different coupling strengths $D$ increasing from left to right. In the second row we show the results for the dissipator $\fd[a_{1} - a_{2}]$ and in the third row for $\fd[a_{1} + a_{2}]$. In all cases $\gamma_{2}/\gamma_{1}=100$ and $\Delta/\gamma_{1}=0.4$.}
\label{fig:WF1disM}
\end{figure*}

In the classical case, the sign of the coupling constant $D$ determines whether the two VdP oscillators synchronize in-phase (positive $D$) or out-of-phase (negative $D$). In the quantum case we have $D>0$ and thus expect in-phase synchronization. However, we can write down a slightly different dissipator, $D \fd[a_{1} + a_{2}] \rho$, which for $D>0$ results in out-of-phase synchronization, as we shall see below.

We note that dissipatively coupled quantum VdP oscillators have recently been studied in Ref.~\cite{Lee2014} with a focus on phase locking and the limit $\gamma_{2}^{(i)}/\gamma_{1}^{(i)} \to \infty$. In this paper, we instead focus on frequency entrainment both in the quantum limit $\gamma_{2}^{(i)}/\gamma_{1}^{(i)} \to \infty$ as well as in the more general case where the quantum VdP oscillators cannot be approximated by two-level systems.

\subsection{Steady-state properties}\label{sec:steadystate}

We start by investigating the steady-state properties of the system by calculating the steady-state density matrix $\rho$, the Wigner function $W(x_{i},p_{i})$ of each oscillator, and the joint-probability distribution $P(x_{1}, x_{2})$ for the positions of the two oscillators.

\subsubsection{Steady-state density matrix}

Let us first consider the case of equal limit cycles, $\gamma_{1}^{(1)} = \gamma_{1}^{(2)} = \gamma_{1}$ and
$\gamma_{2}^{(1)} = \gamma_{2}^{(2)} = \gamma_{2}$. In the quantum limit $\gamma_{2}/\gamma_{1} \to \infty$, an analytical expression for the steady-state density matrix can be obtained~\cite{Lee2014}. For zero coupling $D=0$ the density matrix of the system factorizes, i.e., $\rho = \rho_{1} \otimes \rho_{2}$, where $\rho_{i} = \trm{diag}( 2/3, 1/3 )$ is the steady state of the $i$-th quantum VdP oscillator~\cite{Lee2013, Walter2014}. For finite coupling $D\not= 0$, the off-diagonal matrix elements $\bra{01} \rho \ket{01} = \bra{10} \rho \ket{10}$ and $\bra{10} \rho \ket{01} = \bra{01} \rho \ket{10}^{*}$ become non-zero. In addition, the ground-state occupation $\bra{00} \rho \ket{00}$ increases as the dissipative coupling increases the total damping. However, the two reduced density matrices remain diagonal, i.e.,~the coupling
induces only coherence between the two oscillators and not between the Fock states of one oscillator.

For finite $\gamma_{2}/\gamma_{1}$ the oscillators will occupy Fock states with $n>1$. For a nonzero coupling $D\not=0$ off-diagonal matrix elements other than $\bra{01} \rho \ket{01}$ start to become important and  contribute to finite coherences $\langle a_{1}^{\dagger} a_{2}^{\pd} \rangle \not= 0$.

\subsubsection{Steady-state Wigner function}

The Wigner function of the $i$-th oscillator is given by
\begin{align}
	W(x_{i},p_{i}) = \int \, \frac{dy}{2\pi} e^{-i p_{i} y} \left< x_{i} + \frac{y}{2} \right| \rho_{i} \left| x_{i} - \frac{y}{2} \right> \, ,
\end{align}
where $\rho_{i}$ is the reduced (steady-state) density matrix of oscillator $i$. The Wigner function is a quasi-probability distribution, i.e.,~it can become negative. We also calculate the joint-probability distribution $P(x_{1}, x_{2})$ for the position variables $x_1$ and $x_2$ of the two oscillators
\begin{align}
	P(x_{1}, x_{2}) &= \left< x_{1} , x_{2} \right| \rho \left| x_{1}, x_{2} \right> \,
\end{align}
which is a probability distribution and non-negative.

In Fig.~\ref{fig:WF1disM} we show the Wigner function $W(x_{1},p_{1})$ and the joint-probability distribution $P(x_{1}, x_{2})$ for different coupling strengths $D$. For zero coupling $D=0$ the density matrix of the system factorizes and the joint-probability distribution becomes $P(x_{1}, x_{2}) = \left< x_{1} \right| \rho_{1} \left| x_{1} \right>  \left< x_{2} \right| \rho_{2} \left| x_{2} \right>$. The Wigner function of both oscillators has a ring shape, i.e.,~the system has a finite amplitude while the phase of this oscillation is undetermined. This is reminiscent of a classical limit cycle where the amplitude is fixed but the phase is free. As the coupling increases the weight of the joint-probability distribution $P(x_{1}, x_{2})$ becomes gradually concentrated near the line $x_{1} = x_{2}$. This indicates the build up of correlations between the motion of two VdP oscillators: the stronger the coupling, the more closely the oscillators are moving together. This picture compares well with the phase-space behavior of the classical system: below the synchronization threshold the limit cycle of two coupled VdP oscillators is a ``Lissajous figure'' in the $x_{1}-x_{2}$ plane while the synchronized limit-cycle features a line in the $x_{1}-x_{2}$ plane. We note that the Wigner function $W(x_{i},p_{i})$ remains rotationally symmetric indicating that there is no build-up of coherence between the Fock states of each oscillator and that the reduced density matrix remains diagonal. We also note that as we increase the dissipative coupling $D$, the Wigner function $W(x_{i},p_{i})$ eventually turns from ring shape to a Gaussian centered at the origin due to the additional linear damping.

\subsection{Frequency entrainment}\label{sec:entrainment}

We have already seen that the joint-probability distribution $P(x_{1},x_{2})$ gives a first hint on synchronization. Now we will study the frequencies in the power spectra and see how they behave as the coupling is turned on. This gives us the possibility to characterize synchronization in terms of observed frequencies and their entrainment.

\subsubsection{The spectrum}

%
\begin{figure}
\centering
\includegraphics[width=\columnwidth]{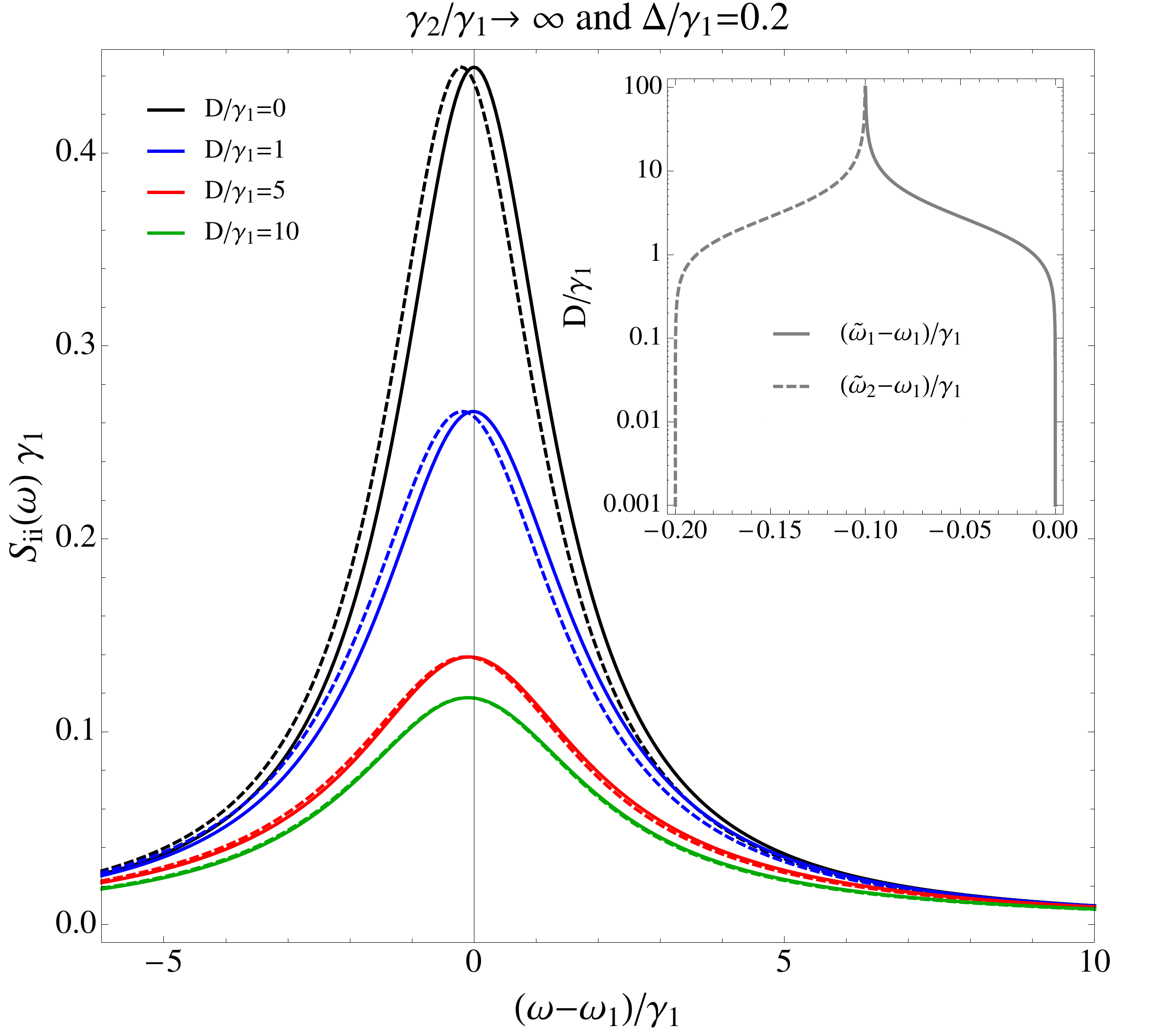}
\caption{ Spectra $S_{11}(\omega)$ (solid) and $S_{22}(\omega)$ (dashed) for different coupling strengths $D$ and for $\gamma_{2}/\gamma_{1} \to \infty$ and $\Delta/\gamma_{1} = 0.2$. Inset: Observed frequencies $\tilde{\omega}_1$ and $\tilde{\omega}_2$ as a function of coupling strength $D$.}
\label{fig:specwf}
\end{figure}

The uncoupled self-sustained VdP oscillators are characterized by their unperturbed frequencies, $\omega_{1}$ and $\omega_{2}$. A non-zero coupling between the two systems affects the frequency of each oscillator. These new frequencies are called observed frequencies denoted $\tilde{\omega}_{1}$ and $\tilde{\omega}_{2}$. With increasing coupling strength $D$ between the two oscillators, the initially different frequencies approach each other.

To investigate this frequency entrainment in the case of two coupled quantum VdP oscillators, we identify the observed frequencies as the frequency of the maximum in the quantum-mechanical power spectrum
\begin{align}
	S_{ii}(\omega) = \int_{-\infty}^{\infty} dt \, e^{i \omega t} \, \langle a_{i}^{\dag}(t) a_{i}^{\pd}(0)\rangle \, ,
\end{align}
where $\langle \cdot \rangle$ denotes the average with respect to the full quantum-mechanical steady-state density matrix.

For equal VdP oscillators $\gamma_{1,2}^{(1)} = \gamma_{1,2}^{(2)} = \gamma_{1,2}$ and in the quantum limit $\gamma_{2}/\gamma_{1} \to \infty$ where only the two lowest Fock states of each oscillator are occupied we can solve for the steady state, cf. Ref.~\cite{Lee2014}. The equations of motions for the density-matrix elements, together with the quantum regression theorem, allow us to calculate the spectra $S_{ii}(\omega)$ analytically (the expression is too lengthy to be displayed here). In the general case of finite $\gamma_{2}/\gamma_{1}$, we resort to a numerical solution of the quantum master equation (\ref{eqn:maD}).

Having obtained an analytical expression for the spectra $S_{ii}(\omega)$, we can investigate the limit $D \to \infty$. We find that in this limit the spectra of the two oscillators become identical and are given by
\begin{align}\label{eqn:specDinf}
	S_{11}^{\infty}(\omega) = S^{\infty}_{22}(\omega) = \frac{1}{8} \frac{\frac{5}{2} \gamma_{1}}{ \left(\frac{5}{2} \gamma_{1} \right)^{2} + \left(\omega + \frac{1}{2}(\omega_{1}+\omega_{2}) \right)^{2} } \, .
\end{align}
This means that for $D \to \infty$ the oscillators approach the mean of their unperturbed frequencies $\frac{1}{2}(\omega_{1}+\omega_{2})$.

In Fig.~\ref{fig:specwf} we show as a generic example the spectra in the limit $\gamma_{2}/\gamma_{1} \to \infty$ for a fixed initial detuning $\Delta/\gamma_{1} = 0.2$ and for different coupling strengths $D$. With increasing coupling the height of the spectral peak decreases and the spectra become slightly broader. The inset in Fig.~\ref{fig:specwf} shows that with increasing coupling strength $D$ the observed frequencies $\tilde{\omega}_1$ and $\tilde{\omega}_2$ approach $\frac{1}{2}(\omega_{1}+\omega_{2})$ symmetrically.

\subsubsection{Frequency entrainment of identical limit-cycle oscillators in the quantum limit}

\begin{figure}
\centering
\includegraphics[width=\columnwidth]{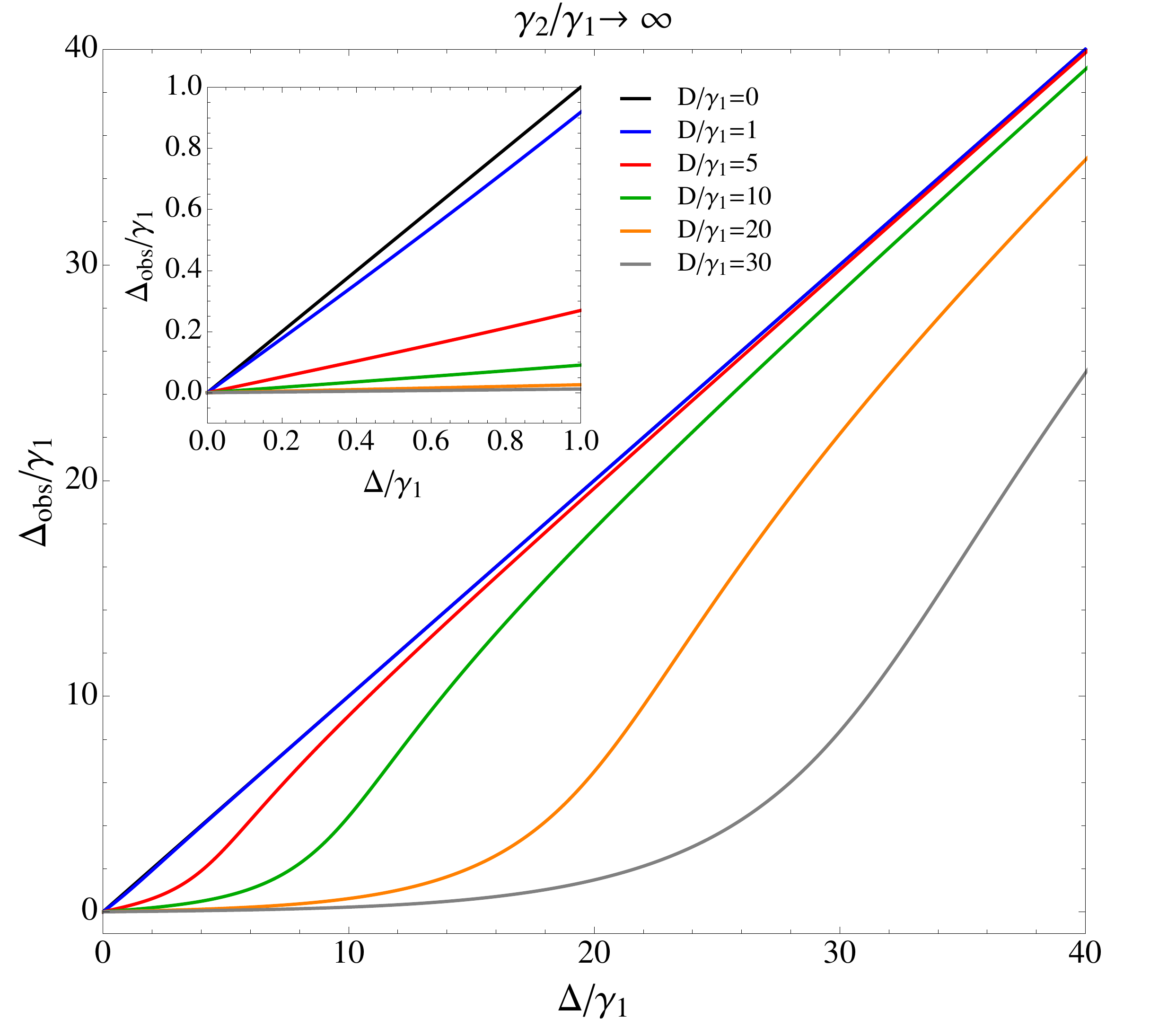}
\caption{\label{fig:adler1}
Observed detuning $\Delta_{\trm{obs}} = |\tilde{\omega}_{2} - \tilde{\omega}_{1}|$ as a function of the initial detuning $\Delta = |\omega_{2} - \omega_{1}|$ for various coupling strengths $D$ and $\gamma_{2}/\gamma_{1} \to \infty$. Increasing the coupling strength leads to frequency entrainment of the two quantum VdP oscillators. A well-defined synchronization plateau as one would expect for classical noiseless  VdP oscillators is absent due to the unavoidable quantum noise.}
\end{figure}

\begin{figure}
\centering
\includegraphics[width=\columnwidth]{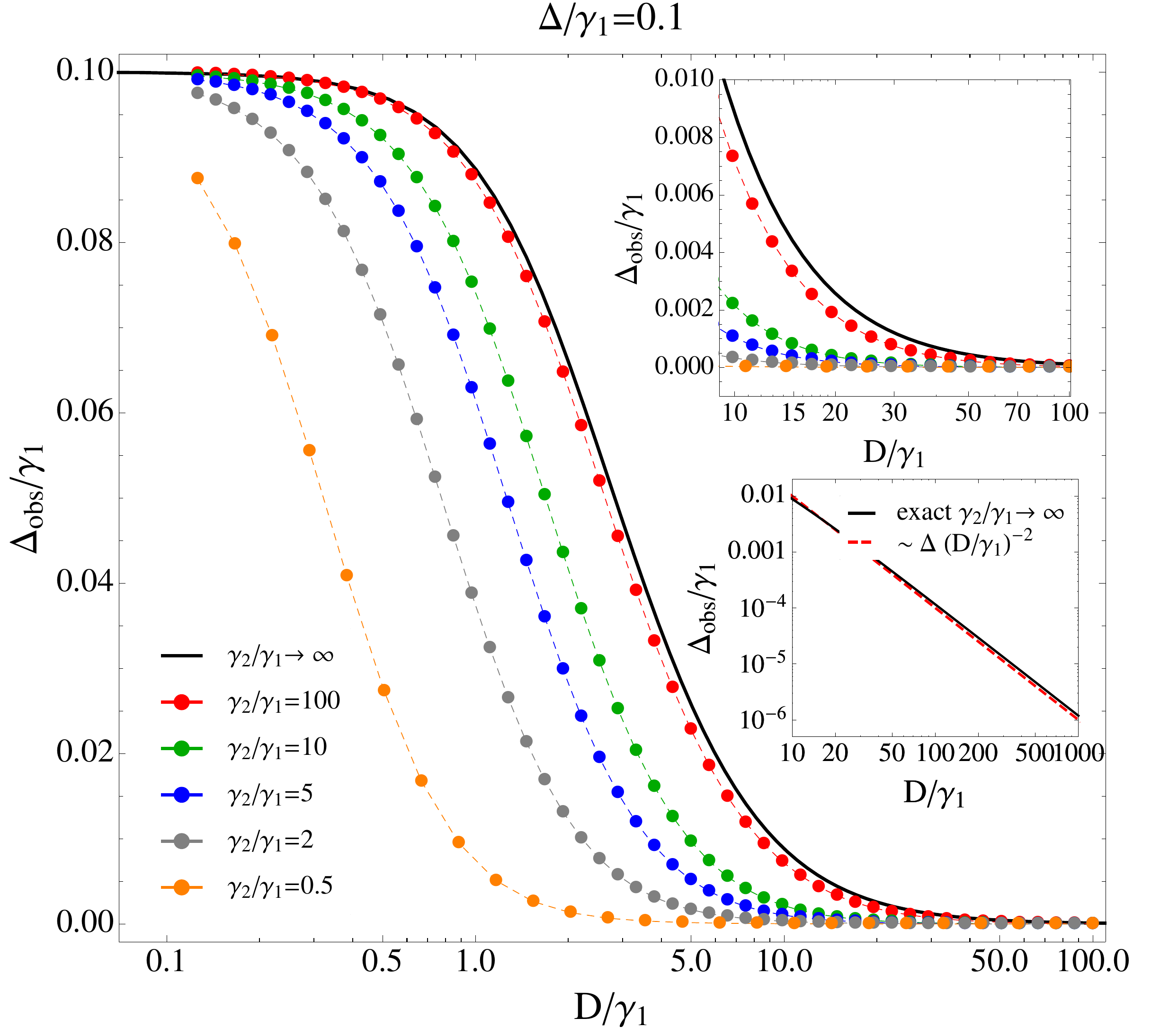}
\caption{\label{fig:DeltaobsVSD2}
Observed detuning $\Delta_{\trm{obs}}$ as a function of the coupling strength $D$ for fixed initial detuning $\Delta/\gamma_{1} = 0.1$ and different $\gamma_{2}/\gamma_{1}$. For increasing coupling $D$ a steplike crossover from weak to strong entrainment is observed. Upper inset: Zoom-in for large $D$.  The entrainment decreases with increasing $\gamma_{2}/\gamma_{1}$ for arbitrary $D$. Lower inset: Observed detuning $\Delta_{\trm{obs}}$ for $\gamma_{2}/\gamma_{1} \to \infty$ goes to zero like $\Delta (D/\gamma_{1})^{-2}$.}
\end{figure}

In Fig.~\ref{fig:adler1} we plot the observed detuning $\Delta_{\trm{obs}} = |\tilde{\omega}_{2} - \tilde{\omega}_{1}|$ over a wide range of values of the initial detuning $\Delta = |\omega_{2} - \omega_{1}|$ for different coupling strengths $D$. Without coupling the observed detuning $\Delta_{\trm{obs}}$ coincides with the initial detuning $\Delta$. Increasing the coupling leads to frequency entrainment. We observe two regimes: for $D \gg \Delta$ we see strong entrainment, while for $D \ll \Delta$ the observed detuning approaches the initial detuning. The inset is a zoom-in at small detunings showing the synchronization plateau of the classical noiseless system is destroyed by quantum noise.

From Fig.~\ref{fig:adler1} we already see that stronger coupling leads to stronger entrainment. To investigate this further, we show in Fig.~\ref{fig:DeltaobsVSD2} the observed detuning $\Delta_{\trm{obs}}$ as a function of the coupling strength $D$ for a fixed detuning $\Delta/\gamma_{1}$ and different values of $\gamma_{2}/\gamma_{1}$. We observe a steplike crossover from weak to strong entrainment with increasing coupling strength $D$. This behavior is similar to what was observed in our previous study of an externally driven quantum VdP oscillator~\cite{Walter2014} where the coupling strength $D$ takes over the role of the external driving strength. In the limit $\gamma_{2}/\gamma_{1} \to \infty$ for large coupling strength $D$, the observed detuning $\Delta_{\trm{obs}}$ vanishes like $\Delta (D/\gamma_{1})^{-2}$, see the lower inset in Fig.~\ref{fig:DeltaobsVSD2}. In contrast to our previous study~\cite{Walter2014}, in the present case larger nonlinear damping rates $\gamma_{2}$ reduce the entrainment above as well as and below the crossover. 

\begin{figure*}
\centering
\includegraphics[width=0.66\columnwidth]{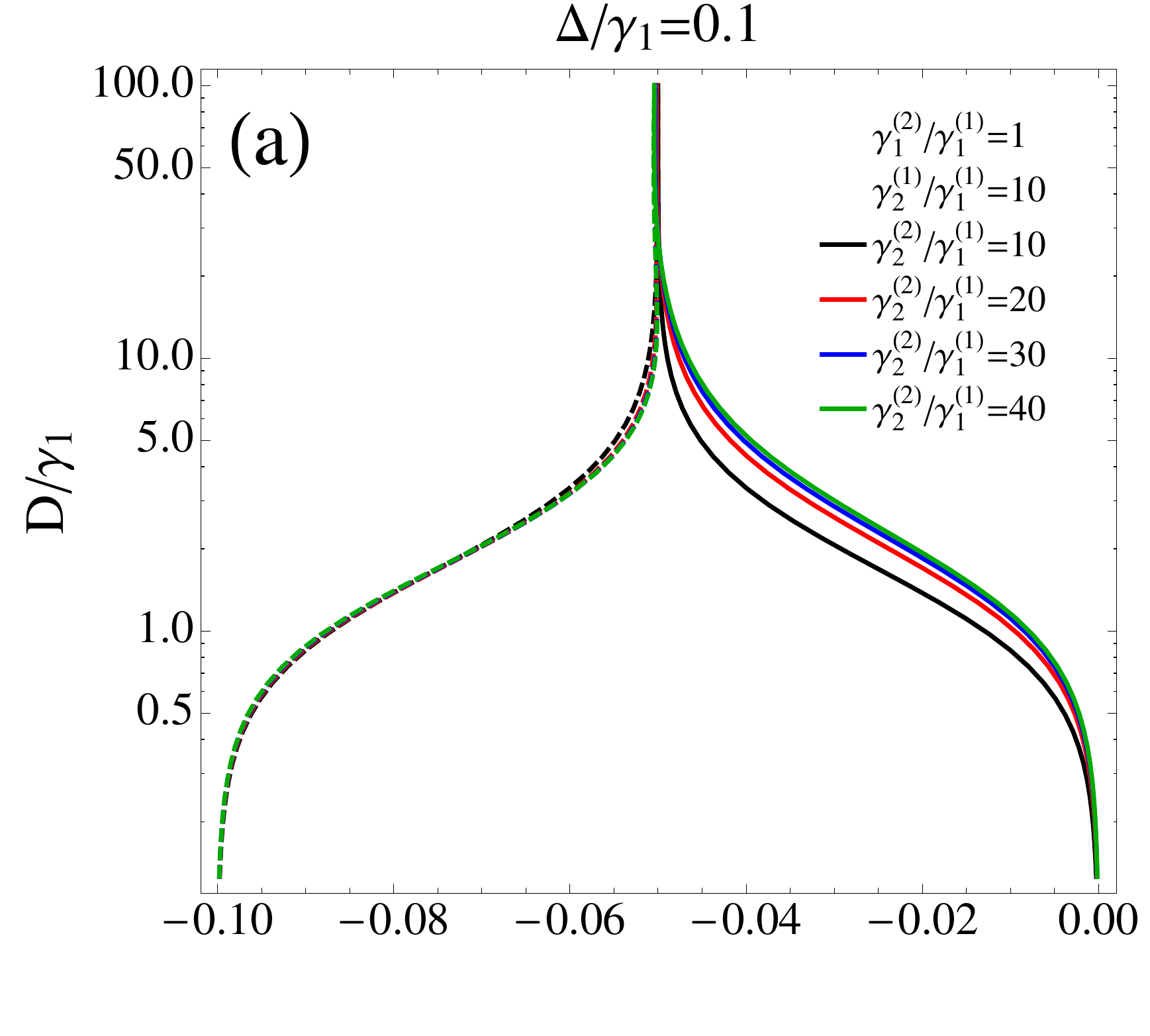}
\includegraphics[width=0.66\columnwidth]{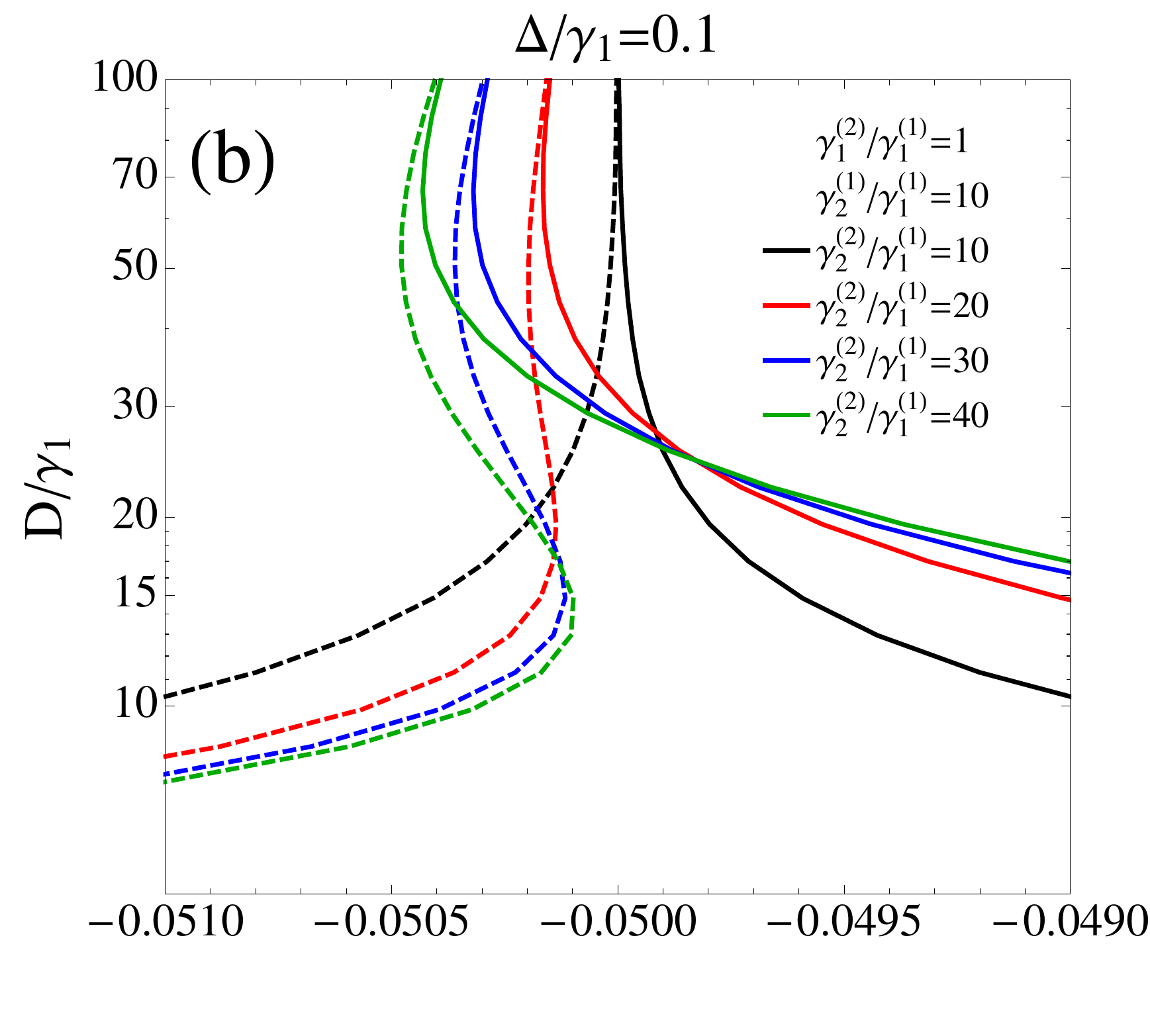}
\includegraphics[width=0.66\columnwidth]{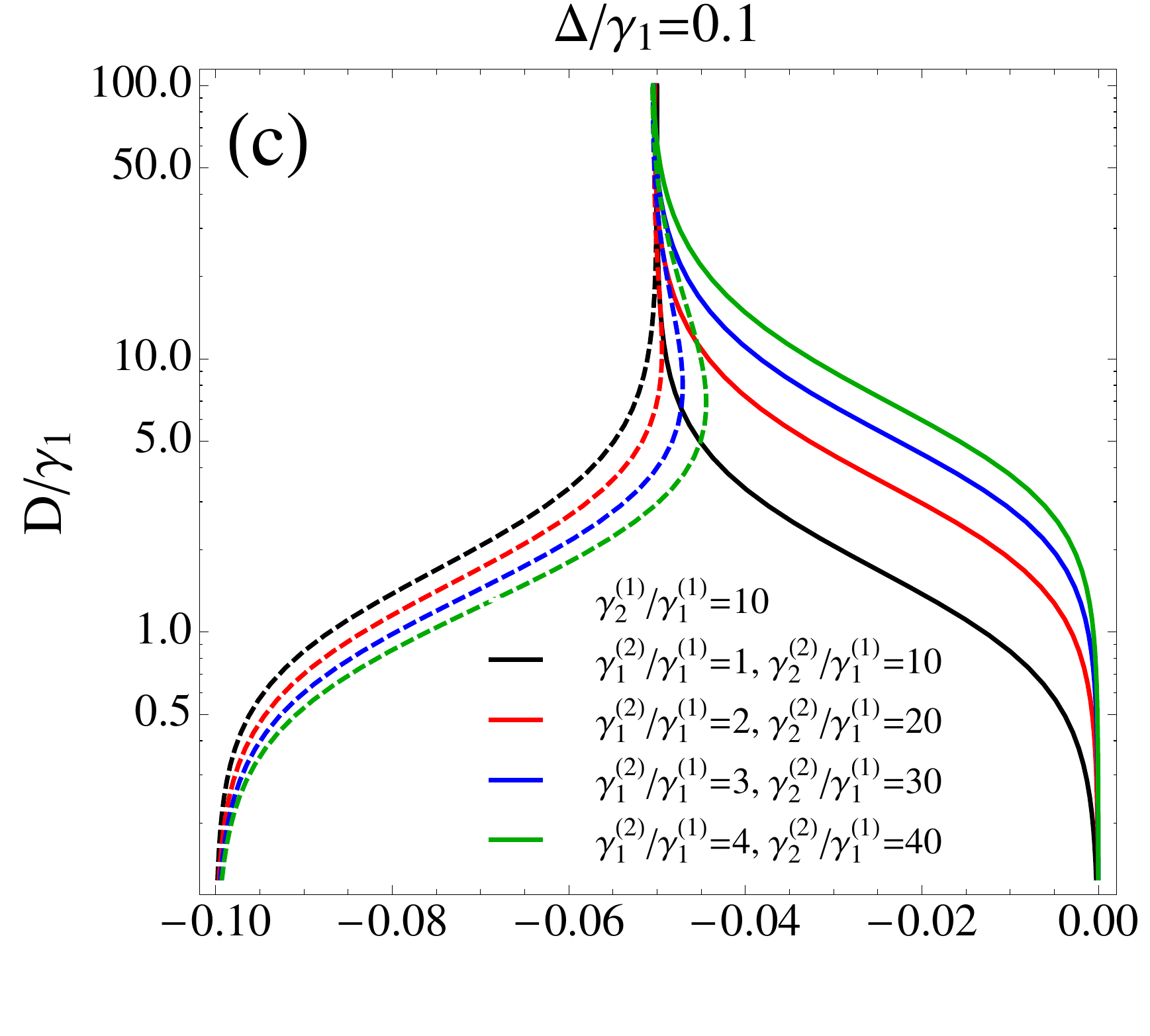}
\caption{\label{fig:wobsOFDuneq1}
Observed frequencies $(\tilde{\omega}_{1}-\omega_1)/\gamma_1$ (full lines) and $(\tilde{\omega}_{2}-\omega_1)/\gamma_1$ (dashed lines) as a function of the coupling strength $D$ for non-identical VdP oscillators: (a) for different limit-cycle radii $\gamma^{(1)}_1/\gamma^{(1)}_2 \not=\gamma^{(2)}_1/\gamma^{(2)}_2$, but equal negative damping parameters $\gamma^{(1)}_1 = \gamma^{(2)}_1$, (b) is a zoom-in for frequencies close to the mean of the unperturbed frequencies, and (c) for equal limit-cycle radii $\gamma^{(1)}_1/\gamma^{(1)}_2 =\gamma^{(2)}_1/\gamma^{(2)}_2$, but different negative damping parameters $\gamma^{(1)}_1 \not= \gamma^{(2)}_1$. The initial detuning in all cases is $\Delta/\gamma^{(1)}_1 = 0.1$.}
\end{figure*}

\subsubsection{Frequency entrainment of non-identical limit-cycle oscillators in the quantum limit}

Up to now, we have restricted our attention to detuned but otherwise identical oscillators. In the following we investigate frequency entrainment between two non-identical quantum VdP oscillators that differ in their damping parameters: (i) the oscillators can have different limit-cycle radii, i.e., $\gamma_{1}^{(1)}/ \gamma_{2}^{(1)} \neq \gamma_{1}^{(2)}/ \gamma_{2}^{(2)}$ or (ii) they can have different negative damping rates $\gamma_1^{(1)} \not= \gamma_1^{(2)}$ determining the relaxation rate to the limit cycle.

In Fig.~\ref{fig:wobsOFDuneq1} we show the observed frequencies $\tilde{\omega}_1$ and $\tilde{\omega}_2$ for non-identical oscillators as a function of the coupling strength $D$. In the limit $D \to \infty$, $\tilde{\omega}_{1}$ and $\tilde{\omega}_{2}$ still approach the mean of the unperturbed frequencies $\frac{1}{2}(\omega_1+\omega_2)$. However, in case (ii) the VdP oscillator with the smaller limit cycle reacts much faster than the oscillator with a larger limit cycle. This can be understood from the classical equations of motion for the phases of the two VdP oscillators
\begin{align}
	\dot{\phi}_{1} &= - \omega_{1} + \frac{D}{2} \frac{r_{2}}{r_{1}} \sin(\phi_{2}-\phi_{1}) \label{eqn:vdpC7a} \ , \\
	\dot{\phi}_{2} &= - \omega_{2} - \frac{D}{2} \frac{r_{1}}{r_{2}} \sin(\phi_{2}-\phi_{1}) \label{eqn:vdpC7b} \, .
\end{align}
For $r_{1} > r_{2}$, the prefactor of the coupling term in Eq.~(\ref{eqn:vdpC7a}) is decreased as compared to the prefactor in Eq.~(\ref{eqn:vdpC7b}). In case of different negative damping rates $\gamma_{1}^{(1)} \not= \gamma_{1}^{(2)}$ but equal limit cycles $\gamma_{1}^{(1)}/\gamma_{2}^{(1)} = \gamma_{1}^{(2)}/\gamma_{2}^{(2)}$ the oscillator with a larger negative damping rate changes its frequency much faster than the oscillator with a smaller negative damping rate.

\section{Experimental realization with a cavity optomechanical system}\label{sec:optomech}

Finally, we propose how to realize dissipatively coupled VdP oscillators with a cavity optomechanical system~\cite{AKM2013}.

In a first step, a single VdP oscillator, i.e.~the two dissipators $\fd[a_{i}^{\dag}]$ and $\fd[a_{i}^{2}]$ in Eq.~(\ref{eqn:ma0}), can be realized within a so-called ``membrane-in-the-middle'' setup~\cite{Thompson2008}, where a movable membrane is placed inside a Fabry-P\'{e}rot cavity. If the membrane is positioned at a node or anti-node of the cavity field, the frequency of the cavity mode is parametrically modulated by the position \emph{squared} of the mechanical oscillator. This quadratic optomechanical coupling can be used to induce two-phonon absorption and emission processes~\cite{Nunnenkamp2010}. Driving the cavity with a laser detuned to the red two-phonon sideband leads to nonlinear damping, i.e., the dissipator $\fd[a_{i}^{2}]$. With linear optomechanical coupling to another cavity mode that is driven by a laser on the blue one-phonon sideband, we can then induce the negative damping, i.e., the dissipator $\fd[a_{i}^{\dag}]$.

In a second step, the two mechanical VdP oscillators can be coupled dissipatively via an additional cavity. (A similar scheme has been proposed in Ref.~\cite{Buchmann2014}.) Starting with the master equation~(\ref{eqn:ma0}) we add a cavity mode $c$ that is coupled linearly to both VdP oscillators and driven with a laser at frequency $\omega_{p}$. The total Hamiltonian is given by
\begin{align}
H_\trm{tot} &= \omega_{c} c^{\dag} c^{\pd}+\eta e^{-i\omega_{p} t} c^{\dag} + \eta^{*} e^{i\omega_{p} t} c \\
			&+ \sum_{i=1,2} \omega^{\pd}_{i} a^{\dag}_{i} a^{\pd}_{i} + g_{i} c^{\dag} c (a^{\pd}_{i} + a^{\dag}_{i})  \nn \, ,
\end{align}
where $\omega_{c}$ is the cavity frequency, $g_{i}$ are the single-photon optomechanical coupling strengths, and $\eta$ the driving strength of the laser. Next, we eliminate the time dependence by going into a frame rotating with the laser frequency, we displace the cavity field, i.e., $c \to \bar{c} + c$ with $\bar{c} =\ex{c}$, and assuming small single-photon optomechanical coupling we neglect nonlinear terms. This gives
\begin{align}
	H_\trm{tot} &= -\Delta_{c} c^{\dag} c^{\pd} + \sum_{i=1,2} \omega^{\pd}_{i} a^{\dag}_{i} a^{\pd}_{i} 
			+ G_{i} (c + c^{\dag}) (a^{\pd}_{i} + a^{\dag}_{i}) \, ,
\end{align}
where we introduced the cavity detuning $\Delta_{c} = \omega_{p} - \omega_{c}$ and the many-photon optomechanical coupling strength $G_{i} = g_{i} \bar{c}$. The density operator $\chi = \rho_{c} \otimes \rho$ of the total system satisfies the following master equation
\begin{align}
	\dot{\chi} = - i \left[ H_\trm{tot}, \chi \right] + \kappa \fd[c]\chi +  \sum_{i=1,2} \gamma_{1}^{(i)} \fd[a_{i}^{\dag}] \chi + \gamma_{2}^{(i)} \fd[a_{i}^{2}] \chi \, ,
\end{align}
where $\kappa$ is the cavity dissipation rate. In the limit $\omega_{i} \gg \kappa \gg G_{i}, \gamma_{1}^{(i)}, \gamma_{2}^{(i)}$ the cavity mode $c$ can be adiabatically eliminated. The resulting reduced master equation reads
\begin{align}\label{eqn:eng}
	\dot{\rho} &= \sum_{i=1,2} - i \left[ \omega^{\pd}_{i} a^{\dag}_{i} a^{\pd}_{i}, \rho \right] + \gamma_{1}^{(i)} \fd[a_{i}^{\dag}] \rho + \gamma_{2}^{(i)} \fd[a_{i}^{2}] \rho \\
			&+ \sum_{i=1,2} G_{i}^{2} \Big[ - \frac{a_{i}^{\dag}a_{i}^{\pd}\rho - a_{i}^{\pd}\rho a_{i}^{\dag} }{\kappa/2 - i(\Delta_{c} + \omega_{i})} +	 \frac{a_{i}^{\pd} \rho a_{i}^{\dag} - \rho a_{i}^{\dag} a_{i}^{\pd} }{\kappa/2 + i(\Delta_{c} + \omega_{i})} \nn \\
			& \hspace{5em }- \frac{a_{i}^{\pd}a_{i}^{\dag}\rho - a_{i}^{\dag}\rho a_{i}^{\pd} }{\kappa/2 - i(\Delta_{c} - \omega_{i})} + \frac{a_{i}^{\dag} \rho a_{i}^{\pd}\rho - \rho a_{i}^{\pd} a_{i}^{\dag} }{\kappa/2 + i(\Delta_{c} - \omega_{i})}   \Big] \nn \\
			&+ G_{1}G_{2} \Big[ - \frac{a_{1}^{\dag}a_{2}^{\pd}\rho - a_{2}^{\pd}\rho a_{1}^{\dag} }{\kappa/2 - i(\Delta_{c} + \omega_{2})} + \frac{a_{1}^{\pd} \rho a_{2}^{\dag} - \rho a_{2}^{\dag} a_{1}^{\pd} }{\kappa/2 + i(\Delta_{c} + \omega_{2})} \nn \\
			& \hspace{5em }- \frac{a_{1}^{\pd}a_{2}^{\dag}\rho - a_{2}^{\dag}\rho a_{1}^{\pd} }{\kappa/2 - i(\Delta_{c} - \omega_{2})} + \frac{a_{1}^{\dag} \rho a_{2}^{\pd}\rho - \rho a_{2}^{\pd} a_{1}^{\dag} }{\kappa/2 + i(\Delta_{c} - \omega_{2})}   \Big] \nn \\
			&+ G_{1}G_{2} \Big[ 1 \leftrightarrow 2 \Big] \nn \, ,
\end{align}
where we dropped the off-resonant terms by employing a rotating wave approximation ($\omega_{1} \approx \omega_{2}$). Adjusting the detuning $\Delta_{c}$ we engineer the desired dissipative coupling between the two mechanical oscillators by choosing $G_{1}=G_{2}$ and $\Delta_{c} \approx -\omega_{1,2}$. The master equation~(\ref{eqn:eng}) then becomes
\begin{align}\label{eqn:eng2}
	\dot{\rho} 	&= \sum_{i=1,2} - i \left[ \omega^{\pd}_{i} a^{\dag}_{i} a^{\pd}_{i}, \rho \right] + \gamma_{1}^{(i)} \fd[a_{i}^{\dag}] \rho + \gamma_{2}^{(i)} \fd[a_{i}^{2}] \rho \\
			&+ \frac{4 G^{2}}{\kappa} \fd[a_{1}+a_{2}] \rho \nn \, .
\end{align}
This corresponds to Eq.~(\ref{eqn:maD}) with $D = 4 G^{2}/\kappa$. Here, the coupling between the VdP oscillators is given by the dissipator $\fd[a_{1}+a_{2}]$ which leads to out-of-phase synchronization, cf.~Fig.~\ref{fig:WF1disM}. The dissipator $\fd[a_{1}-a_{2}]$ can be obtained for $G_{1}=G$ and $G_{2}=-G$.

\section{Conclusion}\label{sec:conclusion}

In this paper, we have investigated synchronization of two dissipatively coupled Van der Pol oscillators in the quantum regime. After a discussion of the classical and quantum equations of motion, we have calculated the steady-state properties of the system. We have used the quantum-mechanical power spectrum to calculate the observed frequencies and the frequency entrainment of the two oscillators. Intrinsic quantum noise prevents strict frequency locking and leads to a crossover from weak to strong frequency entrainment, i.e., approximate synchronization at large coupling strengths. In the extreme quantum limit (nonlinear damping parameter much larger than negative damping parameter) analytical expressions can be obtained. Finally, we have presented a proposal for a possible realization of a synchronization experiment of two dissipatively coupled self-sustained oscillators in an optomechanical setup.

\begin{acknowledgements}
This work was financially supported by the Swiss SNF and the NCCR Quantum Science and Technology.
\end{acknowledgements}




\begin{thebibliography}{99}

\bibitem{Pikovsky2001}
A.\ S.\ Pikovsky, M.\ Rosenblum, and J.\ Kurths,
{\it Synchronization: A Universal Concept in Nonlinear Science}
(Cambridge University Press, New York, 2001).	

\bibitem{Kuramoto1984}
Y.\ Kuramoto, Progr. Theoret. Phys. Suppl. {\bf 79}, 223 (1984).
	
\bibitem{Acebron2005}
J.\ A.\ Acebr\'on, L.\ L.\ Bonilla, C.\ J.\ P\'erez Vicente,
F.\ Ritort, and R.\ Spigler, 
Rev. Mod. Phys. {\bf 77}, 137 (2005).

\bibitem{Shepelyansky2006} 
O.\ V.\ Zhirov and D.\ L.\ Shepelyansky,
Eur. Phys. J. D {\bf 38}, 375 (2006).

\bibitem{Haenggi2006}
I.\ Goychuk, J.\ Casado-Pascual, M.\ Morillo, J.\ Lehmann, and P.\ H\"anggi, 
Phys. Rev. Lett. {\bf 97}, 210601 (2006).

\bibitem{Ludwig2012}
M.\ Ludwig and F.\ Marquardt,
Phys. Rev. Lett. {\bf 111}, 073603 (2013).

\bibitem{Mari2013}	
A.\ Mari, A.\ Farace, N.\ Didier, V.\ Giovannetti, and R.\ Fazio,
Phys. Rev. Lett. {\bf 111}, 103605 (2013).

\bibitem{Lee2012}
T.\ E.\ Lee and M.\ C.\ Cross,
Phys. Rev. A {\bf 88}, 013834 (2013).
 
\bibitem{Hriscu2013}
A.\ M.\ Hriscu and Yu.\ V.\ Nazarov,
Phys. Rev. Lett. {\bf 110}, 097002 (2013).	

\bibitem{Lee2013}
T.\ E.\ Lee and H.\ R.\ Sadeghpour,
Phys. Rev. Lett. {\bf 111}, 234101 (2013).

\bibitem{Holland2013}
M.\ Xu, D.\ A.\ Tieri, E.\ C.\ Fine, J.\ K.\ Thompson, and M.\ J.\ Holland
arXiv:1307.5891 (2013).

\bibitem{Walter2014}
S.\ Walter, A.\ Nunnenkamp, and C.\ Bruder,	
Phys. Rev. Lett. {\bf 112}, 094102 (2014).

\bibitem{Lee2014}
T.\ E.\ Lee, C.\-K.\ Chan, and S.\ Wang,	
Phys. Rev. E {\bf 89}, 022913 (2014).

\bibitem{AKM2013}
M.\ Aspelmeyer, T.\ J.\ Kippenberg, and F.\ Marquardt,
arXiv:1303.0733 (2013).

\bibitem{McEuen2012}
M.\ Zhang, G.\ S.\ Wiederhecker, S.\ Manipatruni, A.\ Barnard,
P.\ McEuen, and M.\ Lipson,
Phys. Rev. Lett. {\bf 109}, 233906 (2012).

\bibitem{Bagheri2013} 
M.\ Bagheri, M.\ Poot, L.\ Fan,  F.\ Marquardt, and H.\ X.\ Tang,		
Phys. Rev. Lett. {\bf 111}, 213902 (2013).

\bibitem{Roukes2014} 
M.\ H.\ Matheny, M.\ Grau, L.\ G.\ Villanueva, R.\ B.\ Karabalin, M.\ C.\ Cross,
and M.\ L.\ Roukes,
Phys. Rev. Lett. {\bf 112}, 014101 (2014).

\bibitem{Kuznetsov2009}
A.\ P.\ Kuznetsov, N.\ V.\ Stankevish, and L.\ V.\ Turukina,
Physica D {\bf 238}, 1203 (2009).

\bibitem{Adler1946}
R.\ Adler,
Proc. IRE {\bf 34}, 351 (1946);
reprinted as Proc. IEEE {\bf 61}, 1380 (1973).

\bibitem{Thompson2008}
J.\ D.\ Thompson, B.\ M.\ Zwickl, A.\ M.\ Jayich, F.\ Marquardt, 
S.\ M.\ Girvin, and J.\ G.\ E.\ Harris,
Nature {\bf 452}, 72 (2008).

\bibitem{Nunnenkamp2010} 
A.\ Nunnenkamp, K.\ B\o{}rkje, J.\ G.\ E.\ Harris, and S.\ M.\ Girvin,
Phys. Rev. A {\bf 82}, 021806(R) (2010).

\bibitem{Buchmann2014}
L.\ F.\ Buchmann and D.\ M.\ Stamper-Kurn,
arXiv:1404.4855 (2014).

\end{thebibliography}
\end{document}